\documentclass[12pt, manuscript, apj]{emulateapj}
\usepackage{amssymb,amsmath}
\usepackage{graphicx}
\usepackage{natbib}

\def\Ha{\rm H\alpha}
\def\Haa{{\rm H\alpha_1}}
\def\Haaa{{\rm H\alpha_2}}
\def\Hb{\rm H\beta}

\def\sHa{\sigma_{\rm H\alpha}}
\def\sHaa{\sigma_{\rm H\alpha_1}}
\def\sHaaa{\sigma_{\rm H\alpha_2}}
\def\sHb{\sigma_{\rm H\beta}}
\def\RHa{R_{\rm BLR}^{\rm H\alpha}}
\def\RHaa{R_{\rm BLR}^{\rm H\alpha_1}}
\def\RHaaa{R_{\rm BLR}^{\rm H\alpha_2}}
\def\RHb{R_{\rm BLR}^{\rm H\beta}}

\def\nii{[N~{\sc ii}]}
\def\oi{[O~{\sc i}]}
\def\oiii{[O~{\sc iii}]}
\def\sii{[S~{\sc ii}]}

\shorttitle{Intermediate BLR}
\shortauthors{Zhang X.-G.}

\begin{document}
\title{Evidence for Intermediate BLR of Reverberation-Mapped AGN PG 0052+251}
\author{Xue-Guang Zhang\altaffilmark{1,2}}
\altaffiltext{1}{Purple Mountain Observatory, Chinese Academy of Sciences,
            2 Beijing Xi Lu, Nanjing, Jiangsu, 210008, P. R. China;
            xgzhang@pmo.ac.cn}
\altaffiltext{2}{Department of Physics and Astronomy, Texas A\&M University,
            College Station, TX 77843-4242, USA}


\begin{abstract}
    In this manuscript, we study properties of BLR of well-known 
reverberation-mapped AGN, in order to find reliable evidence for 
intermediate BLR. We firstly check properties of mapped AGN 
collected from literature in plane of $\sHb^2/\sHa^2$  vs $\RHa/\RHb$. 
Commonly, virial BH masses based on observed broad H$\alpha$ and H$\beta$ 
should be coincident. However, among the mapped objects, PG0052 and NGC4253 
are two apparent outliers in the plane of $\sHb^2/\sHa^2$ vs $\RHa/\RHb$, 
which indicate BLRs of PG0052 and NGC4253 have some special characters. 
Then based on the 55 public spectra of PG0052, BLR of PG0052 is been carefully 
studied in detail.  We find that line width ratio of total observed broad 
H$\alpha$ to total observed broad H$\beta$ is $\sim$0.7, which is much smaller 
than theoretical/observational value of $\sim$0.9. Furthermore, flux ratio of 
total broad H$\alpha$ to total broad H$\beta$ is about 6.8 (Balmer Decrement), 
which is not one reasonable value for BLUE quasar PG 0052+251. Moreover, 
properties of line cores based on PCA technique confirm there is one inner 
broad component and one seriously obscured intermediate broad component in 
BLR of PG0052. If the seriously obscured intermediate BLR was accepted, properties 
of PG0052 in the plane of $\sHb^2/\sHa^2$ vs $\RHa/\RHb$ could be well reproduced, 
which indicates that the intermediate BLR actually is well appropriate to mapped 
quasar PG 0052+251. Finally, the large distance between inner component of BLR and 
intermediate component of BLR  based on CCF results rejects the possibility that the 
intermediate component is probably extended part of inner component of BLR.
\end{abstract}


\section{Introduction}
   Through strong broad emission lines coming from Broad Line Regions (BLR)
of Active Galactic Nuclei (AGN), properties of BLR of AGN which can not be 
directly resolved by current observational technique and instruments have 
been understood deeper and deeper (see reviews of \citealt{su00, gas10, kb11, 
dp10, nm10, ss11, pb11} etc., and references therein). Based on properties of 
BLR of AGN (especially virialized emission line clouds in BLR, \citealt{gas88, 
wan99, pw99, om86, dp09, gas09}), the most convenient method to estimate virial 
BH masses of broad line AGN is proposed (\citealt{v02, on04, pe04, pe10, pb06, 
co06, kb07})
\begin{equation}
M_{BH}\propto V^2\times R_{{\rm BLR}}
\end{equation}
where $V$ represents probale rotating velocity of emission line clouds in 
BLR (line width of broad emission line), $R_{{\rm BLR}}$ means distance between 
BLR and central black hole of AGN (size of BLR) which can be measured by time 
lag between broad line emission and continuum emission from long-period observed 
spectra (\citealt{kas00, pe04, ben06, bw09, dp10}) based on the reverberation 
mapping technique (\citealt{bm82, pe93}). There are so far more than 40 nearby 
broad line objects, of which $R_{{\rm BLR}}$ and virial BH masses have been determined 
(\citealt{pe04, dp10, kas05, bw09, bw10, bn11}). Then one empirical relation has 
been found for $R_{{\rm BLR}}$,  
$R_{{\rm BLR}}\propto L_{5100}^{\sim0.5}\propto L_{line}^{\sim0.5}$, 
where $L_{5100}$ means AGN continuum luminosity at 5100\AA and $L_{line}$ represents 
broad line luminosity (\citealt{kas05, dp10, bw09, wz03, gh10}), and 
virial BH masses can be simply and conveniently estimated from single epoch spectra 
of broad line AGN by line width and continuum luminosity (or broad line luminosity) 
(\citealt{v02, md04, su00, nm10, sr06, sb07, sh08, rh11, nm10, kb07, gh05}, 
Denney et al. 2009a)
\begin{equation}
M_{BH}\propto V^2\times L_{5100}^{\sim0.5}\propto V^2\times L_{line}^{\sim0.5}
\end{equation}

      It is clear that from single epoch spectra of broad line AGN, virial BH masses 
based on different broad emission lines should be coincident. However, some conflicting 
results have been reported. \citet{kb07} have found that virial BH masses based on 
broad lines in UV band and broad lines in optical band are coincident well (some simple 
and more recent results can also be found in \citealt{rh11}). However, \citet{sb07} 
have shown that UV broad line CIV$\lambda1545\AA$ is a poor virial estimator. Recently, 
\citet{ad10} have found the opposite that virial BH mass estimates Based on CIV$\lambda1545\AA$ 
are consistent with those based on the balmer lines. Different virial BH masses based 
on different broad emission lines probably provide some kindly deep information of 
structures of BLR of AGN. In this manuscript, to study properties of virial BH masses 
on different broad emission lines for reverberation-mapped AGN is our main objective. In 
order to find more reliable results, parameters based on the reverberation-mapped AGN 
rather than AGN with single epoch spectra should be firstly considered. For recent 
reported reverberation-mapped AGN, sizes of BLR are based on long-period varied broad 
line emission lines in optical band, especially balmer lines. Thus, in the manuscript, 
we mainly consider the properties of balmer lines (H$\alpha$ and H$\beta$) of 
reverberation mapped AGN.  

    Besides the reverberation mapping technique and virialization method for virial BH 
masses, there are some references which study properties of emission lines regions for 
broad optical balmer lines.  \citet{kg04} have shown that after consideration of 
luminosity dependent responsivities, theoretically estimated size of BLR based on 
H$\alpha$ should be 20\% larger than size of BLR based on H$\beta$, and corresponding  
line width of H$\beta$ should be some larger than that of H$\alpha$. The theoretical 
results can be confirmed by observational results shown in \citet{bw10} (size of BLR on 
H$\alpha$ is some larger than size of BLR on H$\beta$) for mapped objects and in 
\citet{gh05} (line width of broad H$\beta$ is some larger than line width of broad 
H$\alpha$) for pure quasars. Furthermore, one called two-component model of BLR 
has been proposed (\citealt{po07, bo09, zz09, hu08}): one inner region for broad 
wings and one intermediate region for cores, in order to explain complexed broad 
lines of AGN. Although estimated (or measured) parameters (line width and size 
of BLR) of H$\alpha$ and H$\beta$ are some different to some extent, estimated 
virial BH masses based on H$\alpha$ and H$\beta$ should be coincident, such as the 
shown confirmed results in \citet{gh05} for pure quasars.

    It is interesting to check the virial BH masses based on H$\alpha$ and H$\beta$ for 
reported reverberation mapped AGN. In the manuscript, we will find that there are 
some mapped AGN, virial BH masses based on H$\alpha$ are much different from BH masses 
on H$\beta$, which will indicate some special information for structure of emission line 
regions for balmer lines. This manuscript is organized as follows. Section 2 shows the 
results based on parameters (measured line width and size of BLR based on long-period 
variations of line emission and continuum emission) of reported reverberation mapped AGN.  
Section 3 gives the detailed results for blue quasar PG 0052+251 with much different virial 
BH masses on H$\alpha$ and on H$\beta$. Section 4 shows the discussion and conclusion. 
In this manuscript, the cosmological parameters $H_{0}=70{\rm km\cdot s}^{-1}{\rm Mpc}^{-1}$,
$\Omega_{\Lambda}=0.7$ and $\Omega_{m}=0.3$ have been adopted here.

\section{Virial BH Masses on Balmer Lines for Mapped AGN}

     As shown in Introduction, we try to check virial BH masses on H$\alpha$ 
and H$\beta$ for mapped AGN. The reported mapped AGN are mainly included in two 
research groups, AGNWATCH group (public data of 10 AGN can be found in
http://www.astronomy.ohio-state.edu/\~{}agnwatch/) and research group in Wise 
Observatory in Tel Aviv University (public data of 17 PG quasars can be found in
http://wise-obs.tau.ac.il/\~{}shai/PG/, \citealt{kas00}). Corresponding references
about observational techniques and instruments for the two groups can be found 
in the two websites above.

   Before proceeding further, we roughly check results reported in the 
literature (especially \citealt{kas00, kas05, pe04, bw10}) for all the 43 reverberation 
mapped AGN, of which observation data are public or unpublic. Here we mainly compare 
properties of broad H$\alpha$ and broad H$\beta$ (the two most strongest broad 
emission lines in optical band of AGN), including line widths (second moment 
rather than FWHM, which is the best approximation of V in Equation 1 as supposed 
by \citealt{pe04}) and sizes of BLR (measured time lag between line emission and 
continuum emission, not the calculated value from empirical relation of 
$R\propto L^{\sim0.5}$) based on H$\alpha$ ($\RHa$) and on H$\beta$ ($\RHb$), 
which are collected and listed in Table 1. Here, only objects with reliable 
parameters (line width and size of BLR), $P>1.5\times P_{\rm err}$ 
(where $P$ means measured parameter, $P_{\rm err}$ represents corresponding 
uncertainty for the parameter) are collected from the literature. Based 
on the simple criterion, some reverberation-mapped objects should be rejected. For 
example NGC3227, its size of BLR based on H$\beta$ is 
$\RHb \sim 8.2^{+5.1}_{-8.4} {\rm light-days}$, uncertainty of 8.4 is larger than 
measured value 8.2, thus NGC3227 is rejected. We also should note that the famous 
reverberation-mapped object NGC5548 is rejected due to the following reasons. 
On the one hand, in \citet{pe04}, line width of broad H$\alpha$ is not reliable 
due to larger uncertainty. On the other hand, we can find that size of BLR are 
similar, but line width of broad H$\beta$ (both second moment and FWHM) are much 
different in \citet{pe04} from those shown in \citet{bw10}, 
$\sHb\sim2000{\rm km/s}$ and ${\rm FWHM}(\Hb)\sim5800{\rm km/s}$ in \citet{pe04}, 
$\sHb\sim4200{\rm km/s}$ and ${\rm FWHM}(\Hb)\sim12000{\rm km/s}$ in \citet{bw10}. 
Thus, NGC 5548 is not included in our parent sample listed in Table 1. Furthermore, 
there is another object, PG0844, we should note. For PG0844, there are reliable 
parameters of line width and size of BLR for broad H$\alpha$ and broad H$\beta$ 
in \citet{kas00}, however, there is no reliable size of BLR based on H$\beta$ in 
\citet{pe04}. Thus PG 0844 is also rejected. Eventually, there are 16 objects listed 
in Table 1, 9 objects have public spectra (7 PG quasars observed by Wise Observatory 
and 2 objects included in AGNWATCH project), and the other 7 objects have no 
public spectra. Furthermore, we should note listed values of parameters in Table 1 
are values from more recent literature. For example, values for PG0052 can be found in 
Kaspi et al. (2000) and in Peterson et al. (2004). Then the listed values for PG0052 
are collected from Peterson et al. (2004).

\begin{figure*}
\centering\includegraphics[height = 7cm,width = 12cm]{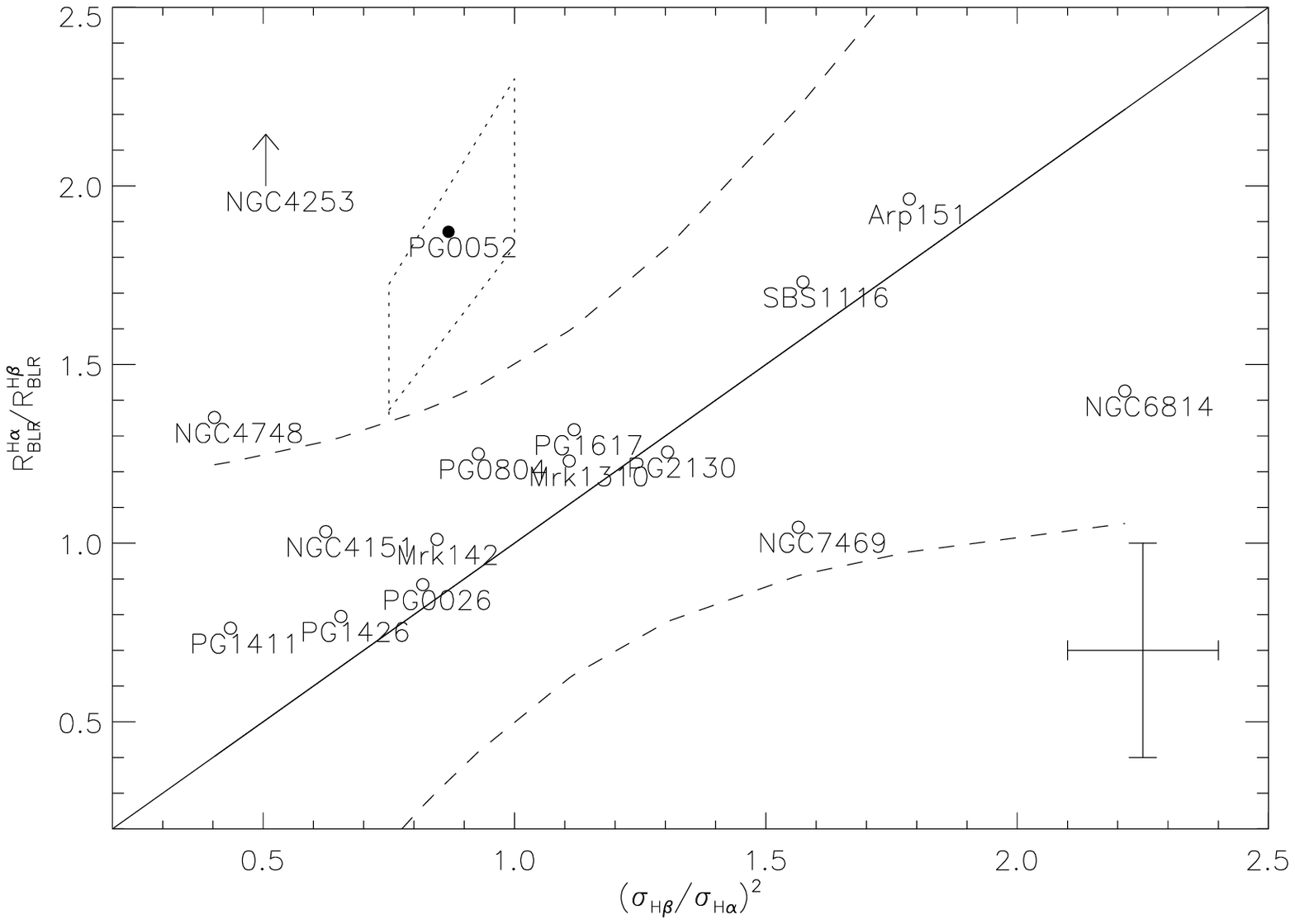}
\caption{On the correlation between line width (second moment) ratio of
$\sHb^2/\sHa^2$ and size ratio of $\RHa/\RHb$. Standard error bar for 
data points is shown in bottom right corner. Object PG 0052+251 is 
marked with solid circle. The area marked by dotted lines around PG 0052+251 
represents the estimated area in the plane, if intermediate BLR was accepted for 
PG0052 (results from equation (6) ). Solid line and dashed line represent
the correlation $\sHb^2/\sHa^2 = \RHa/\RHb$ and its 99\% confidence 
bands. For NGC4253, its true position ([0.51, 4.1]) is much far away from the 
correlation. In order to clearly show the other objects, we use a upper arrow 
to represent that true position of NGC4253 should be much higher than the 
shown position [0.51, 2].}
\label{target}
\end{figure*}

    As shown in introduction, it is clear that there is one strong correlation 
for mapped AGN based on virial BH masses
\begin{equation}
\begin{split}
\sHa^2\times\RHa &= \sHb^2\times\RHb \\
(\frac{\sHa}{\sHb})^2 &= \frac{\RHb}{\RHa}
\end{split}
\end{equation}
In this section, we will check the correlation for the selected mapped AGN. 
Figure~\ref{target} shows the correlation between line width ratio of broad 
H$\beta$ to broad H$\alpha$ ($\sHb^2/\sHa^2$) and size ratio of $\RHa$ to 
$\RHb$ for the 16 reverberation-mapped objects listed in Table 1. For all the 
16 objects, the Spearman Rank correlation coefficient for the correlation is 
about 0.4 with $P_{\rm null}\sim14\%$ (two-sided significance of deviation from 
zero), which indicates there is one rough positive correlation. In the figure, we 
also show the corresponding 99\% confidence bands for the linear correlation 
$\sHb^2/\sHa^2=\RHa/\RHb$.  Based on the results in Figure~\ref{target}, there are 
probable 2 outliers, PG0052 and NGC4253, which are deviating from the linear 
correlation supported by virialization method.  After the two outliers are rejected, 
the re-calculated Spearman Rank correlation coefficient is about 0.9 with 
$P_{\rm null}\sim4\times10^{-5}$.

   Based on the results shown in Figure~\ref{target}, the much different virial 
BH masses based on H$\alpha$ and H$\beta$ should indicate that there are some 
unique characters for broad balmer emission line regions. To find and study the 
character is the main objective of our following section. For the two outliers, 
only PG0052 have public observed spectra which can be downloaded from 
http://wise-obs.tau.ac.il/~shai/PG/.  This, in the following section, we will 
mainly study the properties of reverberation mapped AGN PG0052, and try to 
find some special characters of broad balmer line regions of PG0052.

\section{Main Results for PG0052}

     In this paper, all the observational data and spectra of PG 0052+251
are collected from \citet{kas00} (http://wise-obs.tau.ac.il/\~{}shai/PG/).
The detailed description about the data and spectra can be found in
\citet{kas00}. Here, we do not describe instruments and observational
techniques any more.  There are 56 spectra from 4000$\AA$ to 8000$\AA$ 
with dispersion of $\sim$3.8$\AA$/pixel and spectral resolution about 
$\sim10\AA$ observed from 16th Oct. 1991 to 27th Sep. 1998 for PG 0052+251.  
However, there is one spectrum with many bad pixels around H$\alpha$, thus 
only 55 spectra are considered. The collected spectra from the website above 
have been binned into 1$\AA$ per pixel and been padded from 3000$\AA$ to 
9000$\AA$ as shown in the website. We mainly consider the data and spectra 
as follows.

\subsection{Properties of Mean Spectrum}

    Mean spectrum of PG 0052+251 can be created by PCA (Principal 
Components Analysis or Karhunen-Loeve Transform method, ) technique applied 
for observed noisy spectra.  PCA technique is a mathematical procedure 
that uses an orthogonal transformation to convert a set of observations of 
possibly correlated variables into a set of values of uncorrelated variables 
called principal components. Certainly, mean subtraction (or mean centering) 
is necessary for performing PCA to ensure that the first principal component 
describes the direction of maximum variance. However, if mean subtraction is not 
performed, the first eigencomponent through PCA technique commonly represents 
the mean spectrum of noisy spectra. Here, the convenient and public IDL PCA 
program 'pca\_solve.pro' written by D. Schlegel in Princeton University is used, 
which is included in SDSS software package of IDLSPEC2D (http://spectro.princeton.edu/).

\begin{figure*}
\centering\includegraphics[height = 8cm,width = 14cm]{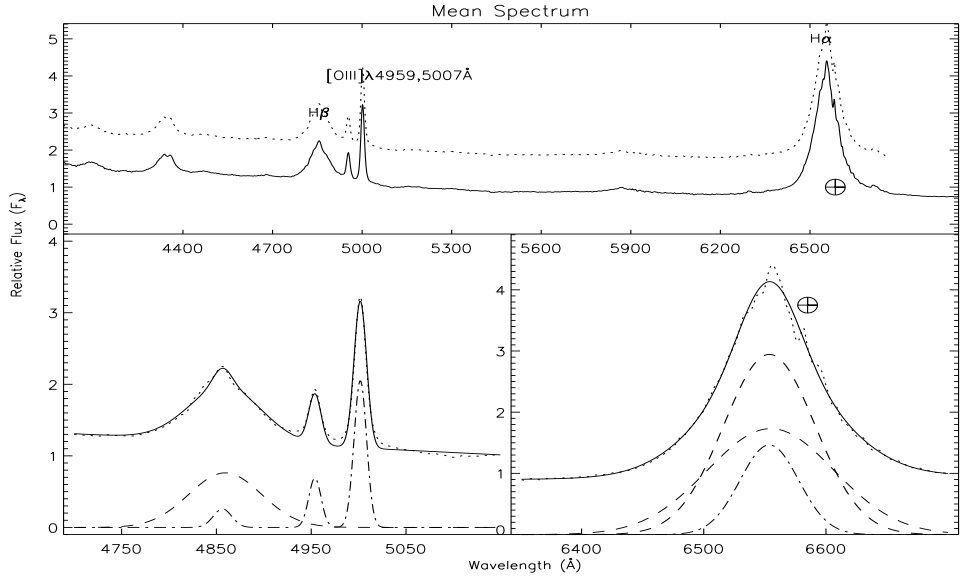}
\caption{The mean spectrum of PG 0052+251, and best fitted results for 
emission lines around H$\alpha$ and H$\beta$. In top panel, solid line 
shows our mean spectrum created by PCA method ($flux(5100\AA)=1$), dotted 
line shows the mean spectrum shown in \citet{kas00} ($flux(5100\AA)=2$). 
In bottom-left panel, dotted line shows the mean spectrum around H$\beta$, 
thick solid line represents the best fitted results, dashed line represents 
broad component of H$\beta$ fitted by one broad gaussian function, dot-dashed 
line shows narrow components of H$\beta$ and [OIII]$\lambda4959,5007\AA$. In 
bottom-right panel, dotted line shows the mean spectrum around H$\alpha$, thick 
solid line represents the best fitted results. Thick dashed line shows broad 
component, if one broad gaussian function is applied to fit H$\alpha$. Thin 
dashed line shows the inner broad component, dot-dashed line shows the intermediate 
broad component, if two broad gaussian functions are applied to fit H$\alpha$. 
Symbols $\oplus$ in the figure show positions of features of atmospheric A band 
near 7620\AA in observed frame.}
\label{spec}
\end{figure*}

   Figure~\ref{spec} shows the mean spectrum of PG 0052+251 with relative flux 
density $flux(5100\AA)=1$. Furthermore, the mean spectrum of PG 0052+251 created 
by \citet{kas00} is also shown in our Figure~\ref{spec}, which is the same as 
our mean spectrum created by PCA method. Then balmer emission lines are fitted 
by simple gaussian functions through Levenberg-Marquardt least-squares minimization 
method. The best fitted results are also shown in Figure~\ref{spec}. 
Emission lines around H$\beta$ can be best fitted by one 
broad gaussian function for broad H$\beta$ and three narrow gaussian functions 
for narrow H$\beta$ and \oiii$\lambda$4959,5007\AA\, doublet. Here when \oiii 
doublet is fitted, we require that \oiii$\lambda$4959, 5007\AA\, have the same 
line width in unit of km/s, and flux ratio of \oiii$\lambda$4959\AA\, to 
\oiii$\lambda$5007\AA\, is theoretical value 0.33 (\citealt{dp07}). Emission lines 
around H$\alpha$ can be fitted by one broad gaussian function. Here we do not 
consider narrow emission lines around H$\alpha$ in the mean spectrum, due to 
much weak \oi$\lambda$6300,6363\AA, \nii$\lambda$6548,6583\AA\, and 
\sii$\lambda$6716,6731\AA\, doublets. Furthermore, we should note that features 
of atmospheric A band near 7620\AA\,  in observed frame can be detected, which 
are marked in Figure~\ref{spec} and in the following Figure~\ref{line} by 
symbol $\oplus$. Here, we simply discuss effects of the features on measured 
line parameters. Line parameters around H$\alpha$ are firstly 
measured without any consideration of effects of the features (functions are 
applied to total observed line profile), and then re-measured with consideration 
of effects of the features (functions are applied to observed line profile with 
features of atmospheric A band rejected). Measured line parameters due to the 
two different procedures are similar. Thus, in the following part, effects of 
features of atmospheric A band near 7620\AA\, are totally ignored. The measured 
line parameters of broad H$\alpha$ and broad H$\beta$ in mean spectrum are 
listed in Table 2.

    Based on the line parameters listed in Table 2, it is clear that line width 
(second moment) of broad H$\beta$ ($\sHb\sim 38$\AA\, $\sim 2370$km/s) is much 
different from line width of broad H$\alpha$ ($\sHa \sim 37$\AA\, $\sim 1700$km/s), 
which is much different from the results shown in \citet{gh05} and in \citet{kg04} 
($\sHb/\sHa\sim1.06 - 1.1$). The results indicate that broad balmr line regions of 
PG0052 probably have some special characters which are some different from common 
quasars. In other words, single broad gaussian function applied to fit broad 
H$\alpha$ and H$\beta$ is not so good. Thus, we consider broad balmer emission 
lines by a different method as follows.  

\subsection{Properties of Line Cores}

\begin{figure*}
\centering\includegraphics[height = 8cm,width = 12cm]{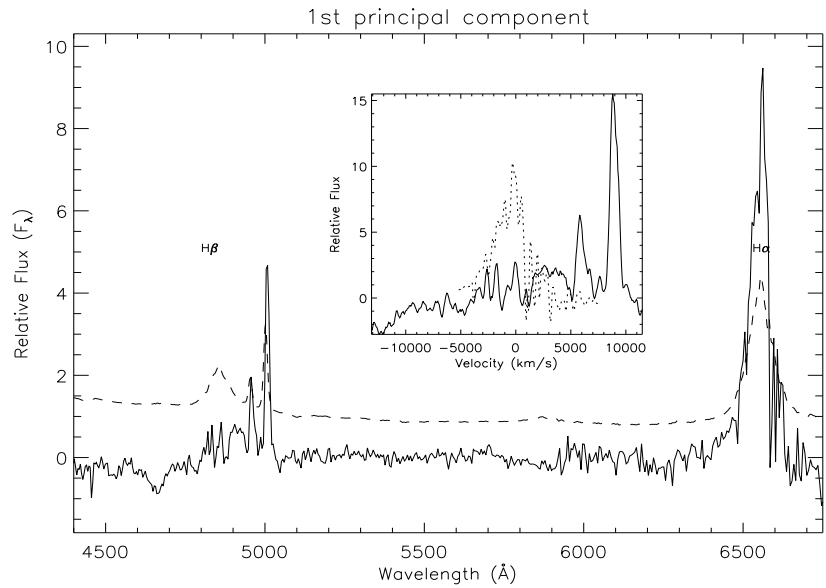}
\caption{ Properties of line cores. Solid line represents the first PCA component 
based on observed spectra with zero mean. Dashed line represents the mean spectrum 
based on observed spectra shown in Figure~\ref{spec}. Middle panel shows the 
comparison of line cores of H$\alpha$ and H$\beta$. In the panel, solid line 
represents line core of H$\beta$ with flux density scaled by 3, dotted line 
represents line core of H$\alpha$.}
\label{pca}
\end{figure*}

    In order to more clearly show properties of broad balmer emission lines, PCA 
technique is applied again as what have been done in \citet{bw94} and in \citet{fh92}. 
Before to study properties of line cores, the commonly accepted step of mean subtraction 
(or mean centering) is firstly performed. Then after the PCA technique applied to the 
spectra with zero mean, the first principal component represents emission line cores 
(\citealt{bw94, fh92}). Figure~\ref{pca} shows emission line cores around H$\beta$ 
and H$\alpha$. In the figure, line profile of H$\beta$ with flux density scaled by 3 
(intrinsic flux ratio of H$\alpha$ to H$\beta$ for blue quasar) is directly compared 
with line profile of H$\alpha$ in the first principal component. Based on the results 
shown in Figure~\ref{pca}, it is clear that the line cores of H$\beta$ and H$\alpha$ 
are much different, there is one apparent broad component in H$\alpha$ but no broad 
component in H$\beta$ in the first principal component. The results indicate besides 
the similar inner broad component as that of observed broad H$\beta$, there is one 
other intermediate broad component in H$\alpha$ but no intermediate broad component 
in H$\beta$ (one seriously obscured intermediate component for balmer lines). 
Moreover, we should note that there are narrow lines around H$\beta$ in the first 
principal component shown in Figure~\ref{pca}, which is due to much extended component of 
[OIII]$\lambda4959,5007\AA$. Similar narrow lines can also be found in rms spectrum of 
PG 0844 shown in Figure 2 in Kaspi et al. (2000). Those narrow lines can not affect our 
results about line cores of H$\alpha$ and H$\beta$.

    Thus, broad H$\alpha$ is fitted again by two gaussian functions, one broad 
gaussian function with similar line width as that of broad H$\beta$ and one 
intermediate broad gaussian function, through Levenberg-Marquardt least-squares 
minimization method. The two components are also shown in bottom-right 
panel in Figure~\ref{spec}. Parameters of the two components of H$\alpha$ are also 
listed in Table 2. The line width (second moment) of intermediate broad component 
of H$\alpha$ is about 1100km/s which is much larger than the line width (second moment)
of narrow \oiii$\lambda$5007\,($\sigma$(\oiii)$\sim6.6$\AA=395km/s). Based on 
flux ratio of inner broad H$\alpha$ to intermediate broad H$\alpha$, $\sim2.6 - 3$, it 
is clear that intermediate broad component of H$\alpha$ is distinct and clearly 
decomposed in the mean spectrum of PG 0052+251. Certainly, we also try to fit broad 
H$\beta$ by two broad gaussian functions. However, intermediate broad component 
of H$\beta$ is not reliable due to smaller measured line width and line 
luminosity than corresponding measured uncertainties for the expected intermediate
component of H$\beta$, which is consistent with results shown in Figure~\ref{pca}. 
The results indicate the intermediate broad component of BLR probably exists and is 
seriously obscured (no intermediate broad component of H$\beta$) for PG 0052+251. 
Certainly, the intermediate broad component of H$\beta$ can not be detected 
perhaps due to low quality of spectra for PG 0052+251. Although there is no true value 
of flux ratio of intermediate broad component of H$\alpha$ to intermediate broad component 
of H$\beta$, it is not difficult to determine cut-off value of S/N (signal to noise) 
for future high quality of spectra. If we accept flux ratio of intermediate broad component 
of H$\alpha$ to intermediate broad component of H$\beta$ is about 6, S/N should be 
larger than 24.

  Before the end of the subsection, we discuss properties of $\chi2$ for fitted 
results for H$\alpha$ and H$\beta$, in order to find more reliable evidence for 
the existence of the second gaussian component for H$\alpha$. Here $\chi2$ is 
value of summed squared residuals for measured parameters divided by degree of 
freedom, which can be used as one good indicator to determine whether fitted results
are acceptable. Here, we accept uncertainty for flux is about $\sim10\%$. The 
values of $\chi^2$ are 0.18, 0.06 and 0.08 for best fitted results by one gaussian functions 
and by two gaussian functions for H$\alpha$ and by gaussian functions for lines around 
H$\beta$ respectively. Then based on fitted results by one or two gaussian functions 
for H$\alpha$, F-test is performed: at 99\% confidence level, calculated variance 
ratio of results by one gaussian function to results by two gaussian functions 
is 2.14 for H$\alpha$ in mean spectrum, which is much larger than F value 1.34. 
The results indicate two gaussian functions for H$\alpha$ is necessary. Due to 
the unreliable second gaussian component for H$\beta$ through least-squares 
minimization method, F-test is not performed for results of H$\beta$.

\subsection{Properties of Observed Spectra}

   We consider observed spectra collected from \citet{kas00} in the subsection. 
Line parameters are measured through gaussian functions for all the 55 observed 
spectra with both H$\alpha$ and H$\beta$ for PG 0052+251, as what we have done 
above for balmer lines in the mean spectrum. Emission lines around H$\beta$ are 
measured once: only a single broad component is fitted for H$\beta$. Emission line 
H$\alpha$ is measured twice, 1): one broad gaussian function is fitted for H$\alpha$, 
2): two broad gaussian functions are fitted for H$\alpha$, one broad component and 
the other intermediate broad component as shown in subsection above. When two gaussian 
functions are applied to fit H$\alpha$, broad component has similar line width as the 
one of broad H$\beta$ (with a permitted scatter of 0.1dex) and line width ratio of 
the corresponding two functions (broad to intermediate broad) is  fixed to 2.16 
($\sim\frac{2370{\rm km/s}}{1100{\rm km/s}}$, based on the measured line widths of two 
components of H$\alpha$ in the mean spectrum). 

    After the measurements of line parameters, we can find if one gaussian function 
is applied to fit H$\alpha$, line width ratio of total broad H$\beta$ to total broad 
H$\alpha$ is about $1.4\pm0.1$, which is much larger than the mean value $\sim$1.1 
for quasars shown in \citet{gh05} and larger than the theoretical value $\sim1.06 - 1.1$ 
in \citet{kg04}. Furthermore, if one gaussian function is applied to fit H$\alpha$, 
flux ratio of total broad H$\alpha$ to total broad H$\beta$ (Balmer decrement) is about 
6.8, which is a larger and unreasonable value for blue quasar PG 0052+251 (mean value 
from composite spectra of quasar is about 3.56 in \citealt{van01}). However, if two 
gaussian functions are applied to fit H$\alpha$, flux ratio of inner broad H$\alpha$ 
to inner broad H$\beta$ (observed total broad H$\beta$ only includes the inner broad 
H$\beta$, intermediate broad H$\beta$ is seriously obscured) is about 3.61 which is 
consistent with the value from composite spectra of quasars (\citealt{van01}), and 
line width ratio is about $\sHa/\sHb\sim0.94$ (here H$\alpha$ is the one without 
intermediate broad H$\alpha$), which is consistent with the result shown in 
\citet{gh05} and in  \citet{kg04}, H$\beta$ is slightly larger than
H$\alpha$ ($\sHa/\sHb\sim0.91$). Certainly, we should note that there are some 
cases that intermediate broad H$\alpha$ is too weak to be detected in observed 
spectra, as one example shown in Figure~\ref{line}, which is probably due to the 
observed spectra with lower resolution. It is clear that total observed broad H$\alpha$ 
separated into two components (one inner broad component and one intermediate 
broad component) should be more reasonable for blue quasar PG 0052+251. 

\subsection{CCF Results}

   Finally, we consider results from cross correlation function (CCF function) 
applied to measure size of BLR of PG 0052 +251, which can be determined by time 
lag between variations of continuum emission and variations of broad lines emission. 
Here the variations of broad emission lines have four components, variations of observed 
broad H$\beta$ (inner broad H$\beta$), variations of observed total broad H$\alpha$, 
variations of inner broad H$\alpha$ (coming from inner BLR) and variations of intermediate 
broad H$\alpha$ (coming from intermediate BLR). Here the used flux densities of total 
broad H$\alpha$, H$\beta$ and continuum emission are collected from \citet{kas00}. 
And then, according to the measured flux ratio of broad H$\alpha$ to intermediate broad 
H$\alpha$ of each observed spectrum, the CORRECTED flux density of total broad H$\alpha$ 
for each observed spectrum can be separated into two values for inner broad component 
and for intermediate broad component of H$\alpha$. Thus, effects from different instruments 
in different configurations can be totally ignored (\citealt{vw92}). In other words, 
the following used flux densities of broad line components are reliable.

     Here, common interpolated cross-correlation function (ICCF) (\citealt{gp86, 
gp87, pe93}) is applied to quantify time lag between continuum emission and broad 
lines emission. We do not consider z-transfer discrete correlation function 
(ZDCF, \citealt{al97, ek88, wp94}) any more, because results from ZDCF are
excellent agreement with results from ICCF (\citealt{pe91, pe92, pe04, kas00, bw10}). 
Figure~\ref{ccf} shows the final results based on the four broad components of 
H$\alpha$ and H$\beta$, including measured sizes of BLR and corresponding 
uncertainties determined by bootstrap method (\citealt{pr92, pe98}). Measured 
sizes of BLR based on variations of different broad components are listed in Table 2. 
The size of inner BLR is about 110 light-days based on inner broad H$\alpha$ (or 
observed broad H$\beta$) and the size of intermediate BLR based on intermediate 
broad H$\alpha$ is about 700 light-days. Furthermore, the measured sizes based on 
observed total broad H$\alpha$ and H$\beta$ are consistent with the reported 
values in \citet{kas00, kas05} and in \citet{pe04}. 

   The clear size of intermediate broad component of H$\alpha$ through CCF results 
indicate the intermediate BLR for H$\alpha$ PG0052 can be mathematically 
determined. Based on the result, we try to discuss some unique characters 
based on intermediate BLR.

\subsection{Intermediate BLR}

   Before to give some clear and further conclusion for mapped AGN PG0052. We 
discuss effects of intermediate BLR on virial BH masses as follows. 

   There are two parameters used for virial BH masses in Equation (1), line width 
and size of BLR. The two parameters are based on observational information. 
If intermediate broad component was accepted for H$\alpha$ of PG0052, we will 
find that virial BH mass estimated from properties of total observed 
broad H$\alpha$ (one inner broad component plus one intermediate broad component)  
should be some different from BH mass estimated from inner (or intermediate) 
broad component, and that estimated virial BH mass based on only 
inner (or intermediate) broad component of H$\alpha$ should be more accurate than 
virial BH mass based on total observed broad H$\alpha$. Under the assumption of 
intermediate BLR accepted for H$\alpha$ of PG0052, we would find that 
\begin{equation} 
\begin{split}
\sHa^2 &\simeq f_1\times\sHaa^2 + f_2\times\sHaaa^2\\
\RHa &\simeq f_1\times\RHaa + f_2\times\RHaaa
\end{split}
\end{equation}
where $\sHaa$ ($\RHaa$) and $\sHaaa$ ($\RHaaa$) mean measured line widths 
(sizes of BLR) based on inner broad component and intermediate broad component 
of H$\alpha$, $f_1$ and $f_2$ represent parameter of flux weight: flux ratio 
of inner (intermediate) broad component to total observed broad H$\alpha$ ($f1+f2=1$). 
One simple but clear method to prove the equation above can be found in Appendix. 

   It is very interesting that for PG0052 (only inner broad component for H$\beta$ but 
inner broad plus intermediate broad component for H$\alpha$), virial BH masses based 
on total observed broad H$\alpha$ should depends on properties of inner and intermediate 
component of H$\alpha$
\begin{equation}
\begin{split}
M_{BH}&\propto\sHa^2\times\RHa\\
       &\propto(f_1^2+f_2^2)(\sHaa^2\times\RHaa)\\
        &+ f_1f_2(\sHaa^2\times\RHaaa+\sHaaa^2\times\RHaa)
\end{split}
\end{equation}
In the equation above, relation $\sHaa^2\times\RHaa\simeq\sHaaa^2\times\RHaaa$ 
is accepted (BH mass estimated through properties of inner broad H$\alpha$ should be 
similar as mass through properties of intermediate broad H$\alpha$). Then we can 
compare BH masses based on observed total broad H$\beta$ (inner broad component for 
H$\beta$) and virial BH masses based on observed total broad H$\alpha$
\begin{equation}
\begin{split}
\frac{\sHa^2\times\RHa}{\sHb^2\times\RHb}\simeq(f_1^2+f_2^2) + 
       f_1f_2(\frac{\RHaa}{\RHaaa}+\frac{\RHaaa}{\RHaa})
\end{split}
\end{equation}
In the equation above, the relation 
$\sHb^2\times\RHb\simeq\sHaa^2\times\RHaa\simeq\sHaaa^2\times\RHaaa$ is accepted 
(BH masses estimated through inner broad H$\beta$ (i.e., observed broad H$\beta$) 
should be similar as mass estimated on inner broad H$\alpha$).
Thus, intermediate broad component of H$\alpha$ for PG0052 should lead to some 
different virial BH mass through total observed broad H$\beta$ (only one inner 
broad component in observed spectrum) and total observed broad H$\alpha$ 
(one inner broad component plus one intermediate broad component), and lead to the 
result that PG0052 should be one outlier in the plane of $\sHb^2/\sHa^2$ vs $\RHa/\RHb$.

   Now let us check results of Equation (6) for PG 0052+251. Flux ratio of inner 
broad H$\alpha$ to intermediate broad H$\alpha$ (total observed broad H$\alpha$ minus 
intermediate broad H$\alpha$) is about 2.5 to 3 (2.5 from mean spectrum, and 3 for 
observed spectra), size ratio of intermediate BLR (for intermediate broad H$\alpha$) 
to inner BLR (for inner broad H$\alpha$) is about 6-7. Then the ratio of 
$\sHa^2\times\RHa$ to $\sHb^2\times \RHb$ should be about 2 based on Equation (6),  
which is well agree with position of PG 0052+251 in the Figure~\ref{target}. In
the Figure~\ref{target}, the area marked by dotted lines shows the accepted range for 
ratio of $\sHa^2\times\RHa$ to $\sHb^2\times \RHb$ under the size ratio of 
$\RHaaa/\RHaa\sim6-7$, based on Equation (6). Furthermore, it is simple to check 
and confirm the correlations shown in Equation (4). The results indicate outlier 
PG 0052+251 in Figure~\ref{target} can be perfectly explained by properties of 
intermediate broad component of H$\alpha$, which further indicate the intermediate 
broad component (or intermediate BLR) is reasonable.

    Before the end of the subsection, we can compare our measured size of intermediate 
BLR with reported results about size of intermediate BLR in literature. Zhu et al. (2009) 
have found one correlation between size of intermediate BLR and central BH masses (Figure 7 
in Zhu et al. 2009). If we accept $M_{BH}\sim(3.69\pm0.76)\times10^8 {\rm M_{\odot}}$ for 
PG 0052+251 (Peterson et al. 2004), then estimated size of intermediate BLR based on correlation 
shown in Zhu et al. (2009) is consistent with our measured value of $\sim10^{18}{\rm cm}$, 
which provide further reliable evidence for intermediate BLR.

\subsection{Extended Part of Inner BLR?}

    One question for existence of intermediate BLR is whether the component is 
only probable extended part of inner BLR, but not one isolated region. We answer the 
question as follows. Based on the results above, there are large distance between inner 
BLR and intermediate BLR, about 600light-days. If the probable intermediate BLR is 
just the extended part of inner BLR, the BLR of PG0052 should have a much extended size. 
So large extended size of BLR should smooth variations of observed broad emission lines, 
in other words, there should be no apparent variations of broad emission lines. We check 
the effects of extended size of BLR on variations by following mathematical procedure. 

    Before proceeding further, we accept the assumptions listed in \citet{pe93} for 
reverberation mapping technique, 1): continuum emission is from one central 
source which is much smaller than BLR, 2): both continuum emission and line emission 
are freely and isotopically propagating in central volume, 3): line emissions 
are in rapid response to ionizing continuum. Furthermore, we accept that line intensity 
from one region simply depends on number density of the region, $I\propto N(r)$ 
(some detailed discussion for emission lines of AGN can be found in \citealt{ne90, of06}).  
Moreover, we simply accept that BLR is spherical shell (center of the sphere at position 
of central black hole) with depth about 600 light-days (inner layer radius is about 90 
light-days, outer layer is about 700 light-days). Then based on the geometrical structure, 
responsed light curve of emission line based on input light curve of continuum 
emission can be created, as what we have done for 3C390.3 (\citealt{zh11}). The spherical 
shell (BLR) is firstly divided into M layers. As long as M is large enough, the effects 
of depth of each layer can be totally ignored. The line intensity from each layer 
(with radius $r$) can be determined by 
\begin{equation}
I(r)\propto N(r)\propto 1/r^{p}
\end{equation}
In order to find more clear effects of extended size of BLR, $p\sim0$ is accepted. 
If  $p$ is much large, the outer part of BLR should have few contributions to 
line emission. Thus, we select $p=0$, there are similar flux strength from clouds 
in inner part of BLR and from clouds in outer part of BLR, which will show more 
clear effects of extended size of BLR on observed light curve of broad line. 
Once one layer meets the ionizing photos from central source, the 
line intensity of the layer is changed immediately as,
\begin{equation}
I_{i}(t) \propto I_{i}(t-1)\times \frac{con_{i}(t)}{con_{i}(t-1)}
\end{equation}
where $t$ is the date, $I_{i}$ and $con_{i}$ mean line intensity from ith layer 
and arriving continuum emission (discrete data series with time separation of 1 day) 
for the layer.  

  There is one point we should note that in our mathematical procedure, the most 
simplest spherical shell geometry is assumed for BLR. Actually the geometry is perhaps 
some different from true structure of BLR of PG 0052+251. As discussed in \citet{gas09, el09, 
ss11, bh10} and reference therein, BLRs of AGN have a flattened distribution and that we 
always view them near pole-on, and BLR structures are very similar in most AGN. Although, 
the applied spherical shell geometry is oversimplified for PG 0052+251, results based on 
the simplified structure can still indicate true effects of extended size of BLR on 
observed light-curve of broad line. Different structures of BLR should lead to some 
different overall trend of observed light-curves of broad lines (such as one example 
shown in \citealt{zh11}), the large extended size of BLR still smooth observed light-curve 
of broad line. As one example to demonstrate effects of extended size of BLR, the 
oversimplified structure of BLR is significantly valid.

   Due to the simple procedure above, responsed light curve of H$\alpha$ based 
on light curve of continuum emission of PG0052 can be created and shown in 
Figure~\ref{var}. It is clear that the responsed light curve of H$\alpha$ is very 
smooth, if extended size of BLR is about 600 light-days,  i.e., if calculated 
intermediate BLR of PG0052 is just extended part of inner BLR. Furthermore, we 
also show one tested light curve of H$\alpha$, if BLR has smaller extended size 
about 30 light-days. For the case with small extended size of BLR, some subtle 
features (arrows in the figure) shown in light curve of continuum emission can be 
reflected in light curve of emission line, however the features can not be 
found in created light curve of emission line for BLR with larger extended size. 
In the subsection, to clearly describe structures of BLR of PG0052 is not our 
objective. Thus, we do not shown further results any more. The results shown in 
Figure~\ref{var} clearly indicate that the intermediate BLR can not be treated as 
extended part of inner BLR for PG0052, otherwise, the observed light curve 
of H$\alpha$ should be much smooth.

\section{Discussions}

  After study of BLR of AGN for more than half one century, 
some information about geometrical structures of BLR has been 
mathematically determined by transfer function in the 
reverberation mapping technique through some special mathematical 
methods (such as Maximum Entropy Method, \citealt{nn86, pe93, pb94, 
ho91, go93, wh94, pw94, kr94, wi95, bw10, bh10, ss11}). Certainly, 
besides the results through mathematical methods, there are some 
statistical results about structures of BLR of AGN (some reviews 
can be found in \citealt{gas09, el09, dr10}), based on 
properties of observed broad broad emission lines, such as 
the proposed model that one much broader component for wing of 
broad emission line and one intermediate broad component for 
core of broad emission line (\citealt{zz09, hu08, su00, bw94, 
br96, ma96, bo09, po07}). However, evidence is not so sufficient 
to confirm intermediate BLR of AGN. In this paper, besides the 
fitted results for broad Balmer emission lines, properties of 
line cores from PCA technique and the measured size of expected 
intermediate BLR are further confirmed evidence for intermediate 
BLR of PG 0052+251: besides one common inner component of BLR 
(inner BLR) with size about 100 light days, there is one other 
seriously obscured intermediate component of BLR (intermediate 
BLR) with size about 700 light days.

    Before proceeding further, we first simply consider one question what geometry
is envisioned for intermediate BLR, where it suffers much more reddening than 
component giving rise to the inner BLR for PG 0052+251? Actually, it is not difficult 
to answer the question. Obscuration for intermediate BLR is not due to dust torus, 
but due to high density dust clouds (radial moving or not) between line-of-sight and 
intermediate BLR. The dust clouds have apparent effects on intermediate BLR, but no 
effects on inner BLR, due to small size of the clouds. The dust clouds can be confirmed 
by the special kind of AGN, AGN with their types changing between type 1 and type 2, 
such as Mrk1018 (\citealt{co86, go90}), NGC 7603 (\citealt{to76}), NGC 2622 and Mrk 609 
(\citealt{go90}) etc.. The large distance between inner BLR and intermediate BLR for 
PG0052 (about 600light-days) ensures enough space for the dust clouds.

   Furthermore, we consider the vitalization assumption (\citealt{gas88, wan99, pw99}) 
for intermediate BLR. As discussed in \citet{bw94} through simple photoionization
model, properties of inner BLR and intermediate BLR are consistent with virialization 
assumption
\begin{equation}
V_1^2\times R_{{\rm BLR,1}} \sim V_2^2\times R_{{\rm BLR,2}}
\end{equation}
$R_{{\rm BLR}}$ means size of BLR (distance between BLR and central
black hole) and $V$ represents the rotating velocity of broad emission
line clouds in corresponding BLR (commonly, line width of broad emission
lines, the second moment). For PG 0052+251, the sizes of two components
of BLR and line widths of inner broad and intermediate broad components have 
been measured above,
\begin{equation}
\left(\frac{\sHaa}{\sHaaa}\right)^{2.0} = 4.7\pm0.4 \simeq \frac{\RHaaa}{\RHaa} = 6\pm1.5
\end{equation}
Here, the mean value $(\sHaa/\sHaaa)$ is calculated by the measured 
line parameters for all observed spectra of PG 0052+251, which is consistent
with the one from mean spectrum of PG 0052+251. The value of $\RHaaa/\RHaa$ is 
calculated by the measured sizes of BLR listed in Table 2. It is clear that the 
result is consistent with what we expect under the virialization method for BLR.

 Certainly, some effects on measured line parameters of mapped AGN should be 
discussed. As discussed in \citet{kas00}, the variable Fe II lines could alter 
line parameter measurements. However, Among the listed PG quasars in Table 1, 
PG 0052 and PG 0026 are the only two objects without apparent optical FeII lines, 
the other 5 PG quasars have strong and apparent FeII emission lines. However, 
only PG 0052 is one outlier in Figure 1. Thus, effects of FeII on measured line 
parameters can be totally ignored. Another question we should discuss is 
size of BLR could change, such as simple discussion in \citet{kas00} and 
detailed study and discussions for well-known mapped AGN NGC5548 (\citealt{ben06, 
ben07, dp10, pe04, wp96}). However, we can find decreasing line width with increasing 
size of BLR, consistent with expected results under virialization assumption. 
Thus, effects of size of BLR changing could not explain the outliers in Figure 1.

    Before the end of the section, we should note that it should be interesting to 
find candidates for AGN with intermediate BLR by the outliers in the plane of 
$\RHa/\RHb$ vs $\sHb^2/\sHa^2$, especially for objects with much larger values of
($\sHa^2\times \RHa)/(\sHb^2\times \RHb)$. Thus, we will expect that mapped object 
NGC 4253 should be strong candidate with intermediate BLR.  
\vspace{8mm}

    Finally, a simple summary is as follows. We first check properties of mapped 
AGN collected from literature in plane of $\sHb^2/\sHa^2$ vs $\RHa/\RHb$. Commonly, 
virial BH masses based on properties of observed broad H$\alpha$ and H$\beta$ 
should be coincident. However, among the mapped objects with measured sizes of BLR 
and line widths (second moment) based on long-period observed broad H$\alpha$ and 
H$\beta$, PG0052 and NGC4253 are two apparent outliers in the plane of 
$\sHb^2/\sHa^2$ vs $\RHa/\RHb$, which indicate BLRs of PG0052 and NGC4253 have 
some special characters. Then based on 55 public spectra of PG0052 (\citealt{kas00}), 
the BLR of PG0052 is been carefully studied in detail.  We find that line width ratio 
of total observed broad H$\alpha$ to total observed broad H$\beta$ is $\sim$0.7, 
which is much smaller than theoretical/observational value of $\sim$0.9 found by 
\citet{kg04} and in \citet{gh05}. Furthermore, the flux ratio of total broad H$\alpha$ 
to total broad H$\beta$ is about 6.8 (Balmer Decrement), which is not one reasonable 
value (mean value of 3.56 in \citealt{van01}) for BLUE quasar PG 0052+251. Moreover, 
properties of line cores based on PCA technique indicate there is one inner broad 
component and one seriously obscured intermediate broad component in BLR. If the 
seriously obscured intermediate BLR was accepted, properties of PG0052 in the plane of 
$\sHb^2/\sHa^2$ vs $\RHa/\RHb$ could be well reproduced, which indicates that the 
intermediate BLR actually is well appropriate to mapped quasar PG 0052+251. Finally, 
the large distance between inner component of BLR and intermediate component of BLR 
based on CCF results (about 600light-days) indicates the intermediate component 
is not extended part of inner component of BLR.

\section*{Acknowledgements}
 We gratefully acknowledge the anonymous referee for giving us constructive 
comments and suggestions to greatly improve our paper.
ZXG gratefully acknowledges the support from  NSFC-11003034, and 
gratefully thanks Dr. Kaspi S. to provide us the available observed spectra
of PG 0052 (http://wise-obs.tau.ac.il/\~{}shai/PG/), and gratefully thanks 
AGNWATCH Project, and gratefully thanks Prof. Wang J.-M. for the discussions 
and constructive suggestions for the manuscript. This research has made 
use of NASA/IPAC Extragalactic Database (NED) which is operated by the Jet 
Propulsion Laboratory, California Institute of Technology, under contract 
with the National Aeronautics and Space Administration,

\appendix
\section{To prove Equation (4)}

    Based on definition of second moment (\citealt{pe04})
\begin{equation}
\sigma^2 = \frac{\int\lambda^2\times P_{\lambda} d\lambda}{\int P_{\lambda}d\lambda} -
           \left(\frac{\int\lambda P_{\lambda}d\lambda}{\int P_{\lambda}d\lambda}\right)^2 \\
         = \frac{\int\lambda^2\times P_{\lambda} d\lambda}{\int P_{\lambda}d\lambda} - \lambda_0^2
\end{equation}
where function $P$ represents line profile, and $\lambda$ represents wavelength. 
For broad H$\alpha$ of PG0052 which includes two components of H$\alpha_1$ (inner broad 
component) and H$\alpha_2$ (intermediate broad component), we can find the correlation 
between line width of total broad $\Ha$ ($\sHa$) and line widths of the two components 
of $\Ha$ ($\sHaa$ and $\sHaaa$)
\begin{equation}
\begin{split}
\sHa^2 = &f_1\times\sHaa^2+ f_2\times\sHaaa^2 +f_1\times\lambda_0^2(\Haa) \\
         &+ f_2\times\lambda_0^2(\Haaa) -\left[f_1\times\lambda_0(\Haa) +f_2\times\lambda_0(\Haaa)\right]^2\\
       = & f_1\times\sHaa^2+ f_2\times\sHaaa^2 + G[\lambda_0(\Haa), \lambda_0(\Haaa)]
\end{split}
\end{equation}
where $f_1$ and $f_2$ are flux ratios of separated broad components of H$\alpha$ to
total broad H$\alpha$, i.e., $f_1 = \int P_{\lambda, 1}d\lambda/\int P_{\lambda}d\lambda$,
$f_2 = \int P_{\lambda, 2}d\lambda/\int P_{\lambda}d\lambda$ and $f_1+f_2=1$.
$\lambda_0({\rm H\alpha_1})$ and $\lambda_0({\rm H\alpha_2})$ are the first moments 
(center wavelengths) of the two components of broad H$\alpha$. It is clear that if 
inner component and intermediate broad component have not so much large different 
center wavelengths ($\lambda_0(\Haa)\simeq\lambda_0(\Haaa)$), the equation 
$\sHa^2\simeq f_1\times\sHaa^2+ f_2\times\sHaaa^2$ can be safely accepted.

    To mathematically prove equation $\RHa\simeq f_1\times\RHaa + f_2\times\RHaaa$ 
should be very difficult. Here, we prove the equation by following Monte-Carlo method, 
based on completely HOMOGENEOUS light-curves of continuum emission ($C(t)$) and 
broad line emission ($L(t)$) of well-known reverberation-mapped AGN NGC5548 (\citealt{pb02}) 
collected from AGNWATCH project (http://www.astronomy.ohio-state.edu/\~{}agnwatch/).
Light curves of inner and intermediate broad H$\alpha$ are created through 
bootstrap method (one common used monte-carlo method to create a mock population from
a given sample of data) applied to light curve of broad line H$\beta$ of NGC5548.
Observed light curve of NGC5548 including N data points can be described as 
[$t_i$, $flux_i$], where index $i=1,2,...,N$ means the ith data points in light curve,
$t$ and $flux$ represent the observational date and corresponding flux density of 
emission line. In order to simply our mathematic procedure, the date $t_i$ is re-created 
as isolated integral values with step of 1 day based on observed data series of NGC5548. 
Then mock light curves of inner and intermediate broad components of H$\alpha$ can
be created by the following two steps. On the first step, through the bootstrap method,
a new sample of index from 1 to N, $k=1,...N$, is created. Certainly, there are some same
values in sample of k. On the second step, mock light curve of inner broad H$\alpha$ is
created by [$t_i$, $flux_k$], mock light curve of intermediate broad H$\alpha$ is created
by [$t_i+\Delta$, $flux_k\times f_{sca}$], where $\Delta\in[2,160]$ days are integral values 
and represent distances between inner BLR and intermediate BLR, and $f_{sca}\in[0.2,5]$ 
represents flux density ratio of intermediate broad H$\alpha$ to inner broad H$\alpha$. 
Values of $f_{sca}$ ensure both the two broad components of H$\alpha$ are apparent. 
Parameter $\Delta$ ensures that there is longer distance between intermediate BLR and 
central black hole than the distance between inner BLR and central black hole. Then the 
two steps above are repeated, until there are enough mock light curves. Furthermore, based 
on the bootstrap method, effects of simply different geometrical structures of  
inner and intermediate components of BLR have been simply included (simple results 
about effects of geometric structures of BLR on measured size of  BLR can be found in 
\citealt{zz09}). Based on the created mock light curves (date and flux density: 
[$t_{i, inner}$, $flux_{k, inner}$] and [$t_{i, inter}$,$flux_{k, inter}$]) of inner 
and intermediate component, it is easy to create the mock light curves 
([$t_{i,tot}$,$flux_{k,tot}$]) of total broad line (inner component plus intermediate 
component) by 
\begin{equation}
\begin{split}
&t_{i,tot} = t_{i,inter} \ \ if\ \ \ t_{i,inter} = t_{i,inner} \\
&t_{i,tot} \ \ \ rejected \ \ if\ \ \ t_{i,inter} \neq t_{i,inner} \\
&flux_{k,tot} = flux_{k,inner} +flux_{k,tot}\ \  if\ \ \ t_{i,inter} = t_{i,inner}\\
&flux_{k,tot} \ \ \ rejected \ \ if\ \ \ t_{i,inter}\neq t_{i,inner}
\end{split}
\end{equation} 
Then values of $\RHa$, $\RHaa$ and $\RHaaa$ can be determined by CCF function. 
Here we notice that only CCF result with one apparent peak is accepted.  
Figure~\ref{hsize} shows the correlation (coefficient 0.97 with significance 
of deviation from zero $P_{\rm null}\sim0$ for 664 simulated data points) between 
$\RHa$ and $\RHaa\times f_1 + \RHaaa\times f_2$ ($f_1 = 1/(1+f_{\rm sca})$ and 
$f_2 = f_{\rm sca}/(1+f_{\rm sca})$). It is clear that the expected relation about 
size of BLR in Equation (4) can be commonly accepted. Furthermore, the results shown 
in Figure~\ref{hsize} indicate even there are two apparent two components in 
light curve of broad line, the CCF could be single-peaked not double-peaked appearance.

\begin{table*}
\centering
\begin{minipage}{130mm}
\caption{Parameters for the 16 mapped AGN}
\begin{tabular}{lcccccc}
\hline
name & $\sHa$ & $\RHa$ & $\sHb$ & $\RHb$ & Public \\
     & ${\rm km/s}$       & light-days           & ${\rm km/s}$       & light-days & \\
\hline
PG0026+129 & 1961$\pm$135 & 98.1$^{+28.3}_{-25.5}$ & 1773$\pm$285 & 111.0$^{+24.1}_{-28.3}$ & Kaspi \\
PG0052+251 & 1913$\pm$85  & 163.7$^{+58.5}_{-38.3}$ & 1783$\pm$86 & 89.8$^{+24.5}_{-24.1}$  & Kaspi \\
PG0804+761 & 2046$\pm$138 & 183.6$^{+15.3}_{-13.3}$ & 1971$\pm$105 & 146.9$^{+18.8}_{-18.9}$ & Kaspi \\
NGC4151 & 2422$\pm$79 & 3.2$^{+1.9}_{-1.7}$ & 1914$\pm$42 & 3.1$\pm$1.3 & AGNWATCH \\
PG1411+442 & 2437$\pm$196 & 94.7$^{+36.0}_{-31.5}$ & 1607$\pm$169 & 124.3$^{+61.0}_{-61.7}$ & Kaspi \\
PG1426+015 & 4254$\pm$290 & 75.5$^{+30.5}_{-32.5}$ & 3442$\pm$308 & 95.0$^{+29.9}_{-37.1}$ & Kaspi \\
PG1617+175 & 2483$\pm$160 & 94.2$^{+19.1}_{-25.2}$ & 2626$\pm$211 & 71.5$^{+29.6}_{-33.7}$ & Kaspi \\
PG2130+099 & 1421$\pm$80 & 198.4$^{+32.6}_{-23.4}$ & 1623$\pm$86 & 158.1$^{+29.8}_{-18.7}$ & Kaspi \\
NGC7469 & 1164$\pm$68 & 4.7$^{+1.6}_{-1.3}$ & 1456$\pm$207 & 4.5$^{+0.7}_{-0.8}$ & AGNWATCH \\
Mrk142 & 934$\pm$61 & 2.90$^{+1.22}_{-0.92}$ & 859$\pm$102 & 2.87$^{+0.76}_{-0.87}$ & NO\\
SBS1116+583A & 1218$^{+147}_{-99}$ & 4.12$^{+1.41}_{-0.98}$& 1528$\pm$184 & 2.38$^{+0.64}_{-0.51}$& NO \\
Arp151 & 937$\pm$34 & 8.01$^{+1.05}_{-1.00}$& 1252$\pm$46 & 4.08$^{+0.50}_{-0.69}$& NO \\
Mrk1310 & 717$\pm$75 & 4.60$^{+0.67}_{-0.62}$ & 755$\pm$138 & 3.74$^{+0.60}_{-0.62}$& NO \\
NGC4253 & 726$\pm$35 & 25.50$^{+0.66}_{-0.86}$ & 516$\pm$218 & 6.24$^{+1.65}_{-1.24}$ & NO \\
NGC4748 & 1035$\pm$74 & 7.61$^{+3.01}_{-4.64}$ & 657$\pm$91 & 5.63$^{+1.64}_{-2.25}$& NO \\
NGC6814 & 1082$\pm$52 & 9.51$^{+1.91}_{-1.56}$& 1610$\pm$108 & 6.67$^{+0.88}_{-0.90}$& NO\\
\hline
\end{tabular}
\\
Notice:-- 'Kaspi' means public observed spectra can be found from website
http://wise-obs.tau.ac.il/\~{}shai/PG/. \\
'AGNWATCH' mens public observed spectra can be found from the AGNWATCH 
project http://www.astronomy.ohio-state.edu/\~{}agnwatch/.\\
'NO' means there are no public observed spectra, the listed parameters are collected
from Bentz et al. (2010). \\
Size of BLR based on broad Balmer emission lines are the ones determined by the
$\tau_{cent}$, and line widths are the ones measured from rms spectra as discussed in
Peterson et al. (2004).\\ 
We should note: listed values of parameters are values from more recent literature.
\end{minipage}
\end{table*}

\begin{table*}
\centering
\begin{minipage}{130mm}
\caption{Parameters of PG 0052+251}
\begin{tabular}{lccccc}
\hline
Line &  $\sigma$  & flux & $R_{{\rm BLR}}(p)$ & $R_{{\rm BLR}}(c)$ & $R_{{\rm CCF}}(max)$\\
     &  $\AA$ & $10^{-16}{\rm erg/s/cm^2}$ & light-days & light-days & \\
\hline
H$\beta$(broad) & 38.4 & 1426.3 & 109$\pm$25 & 111$\pm$15 & 0.76 \\
H$\alpha$(tot broad) & 37.4 & 9612.5 & 191$\pm$24 & 214$\pm$19 & 0.72 \\
H$\alpha$(inner broad) & 51.8 & 5067.9 & 110$\pm$20 & 116$\pm$16 & 0.68 \\
H$\alpha$(int broad) & 24.1 & 1649.5$^\star$ & 703$\pm$45 & 678$\pm$40 & 0.38 \\
\hline
\end{tabular}
\\
Notice:-- 
H$\alpha$(tot broad) means observed total broad H$\alpha$. H$\alpha$(inner broad)
means inner broad component of H$\alpha$. H$\alpha$(int broad)
represents intermediate broad component of H$\alpha$. \\
Second column gives the second moment of broad component in unit of $\AA$ shown
in mean spectrum of PG 0052+251, third column shows measured mean flux density of 
broad component for 55 observed spectra of PG 0052+251, fourth column shows measured size 
of BLR through  peak value of CCF results, fifth column shows the measured size of BLR 
through central value of CCF results, sixth column shows the maximum coefficient 
through CCF.\\ 
Corresponding error of measured size of BLR is determined through
bootstrap method as shown in Figure~\ref{ccf}\\
1649.5$^\star$ means the mean value is not so reliable and probably lower
than the internal value, due to much weak intermediate broad H$\alpha$ in
some observed spectra.
\end{minipage}
\end{table*}

\bibliographystyle{apj}

\clearpage

\begin{figure*}
\centering\includegraphics[height = 8cm,width = 12cm]{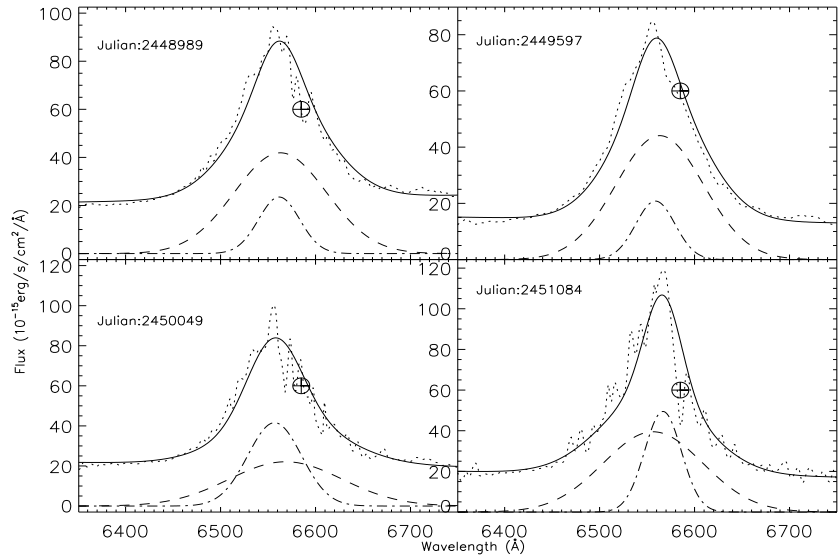}
\caption{Best fitted results for observed H$\alpha$. In each panel, dotted line
shows observed spectrum, thick solid line represents the best fitted
results, dashed line represents inner broad component of H$\alpha$, dot-dashed
line shows intermediate broad component of H$\alpha$. In the case of
Julian:2449596 (top-right panel), there is no apparent intermediate broad component.
Symbols $\oplus$ in the figure show positions of features of atmospheric 
A band near 7620\AA in observed frame.}
\label{line}
\end{figure*}

\begin{figure*}
\centering\includegraphics[height = 14.5cm,width = 16cm]{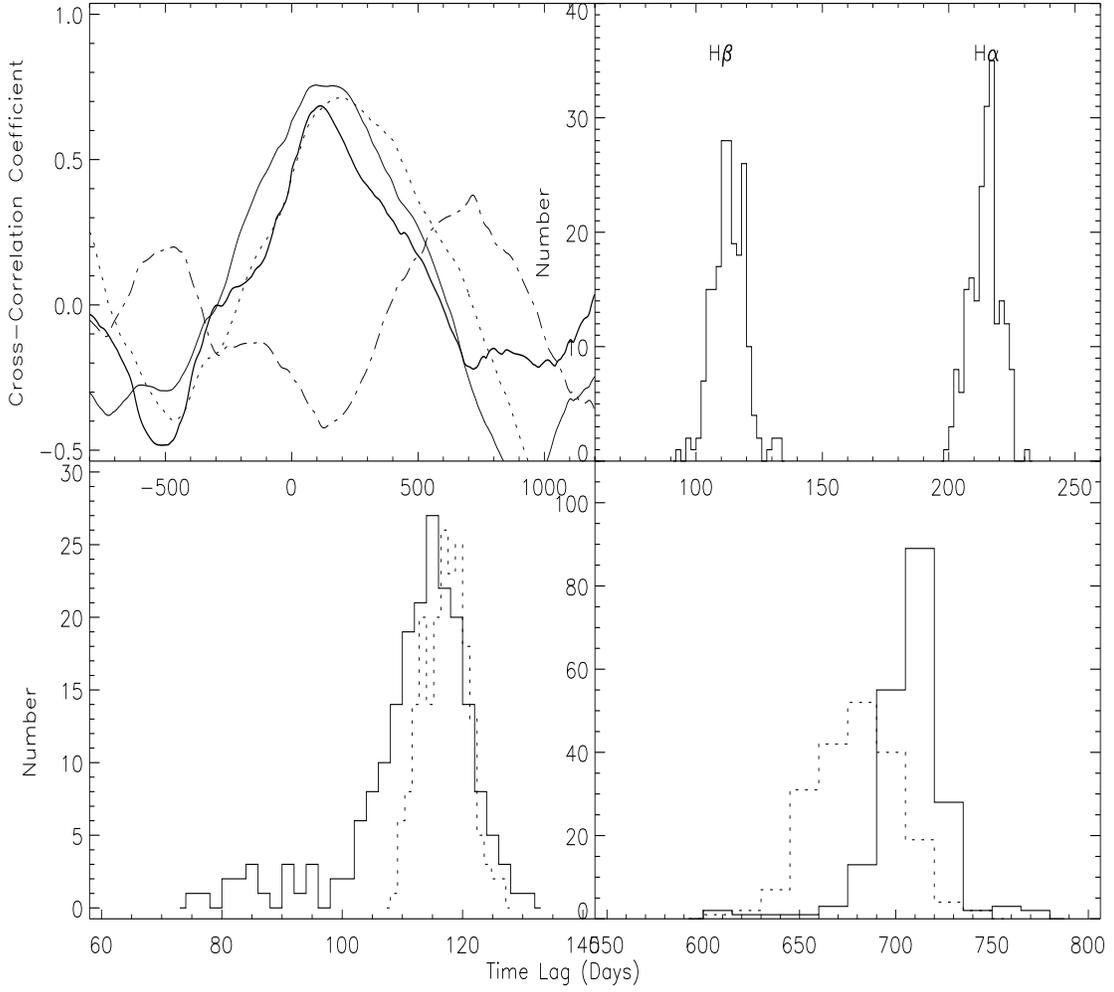}
\caption{Top-left panel shows results about Cross-Correlation Function (CCF). 
Thin solid line represents the result for broad H$\beta$ 
(${\rm CCF}({\rm H}\beta, con_{5100\AA})$). Dotted line represents the result for
total broad H$\alpha$ (${\rm CCF}({\rm H}\alpha_{tot}, con_{5100\AA})$). Thick solid
line represents the result for inner broad H$\alpha$
(${\rm CCF}({\rm H}\alpha_{inner\ broad}, con_{5100\AA})$). Dot-dashed line shows the
result for intermediate broad H$\alpha$
(${\rm CCF}({\rm H}\alpha_{int\ broad}, con_{5100\AA})$). Top right panel shows 
distributions of time-lags between total broad Balmer emission and
AGN continuum emission through bootstrap method, in order to compare
our results with the results shown in Kaspi et al. (2000) and in
Peterson et al. (2004). Bottom-left panel shows distributions of time-lag
between inner broad H$\alpha$ emission (after the subtraction of intermediate
broad H$\alpha$) and AGN continuum emission through bootstrap method.
Bottom-right panel shows distributions of time-lag between intermediate
broad H$\alpha$ emission and AGN continuum emission through bootstrap method.
In the two bottom panels, solid line represents the distributions for time-lag
determined by peak values of CCF results, dotted line represents the distributions
of time-lag determined by central values of CCF results. }
\label{ccf}
\end{figure*}

\begin{figure*}
\centering\includegraphics[height = 10cm,width = 14cm]{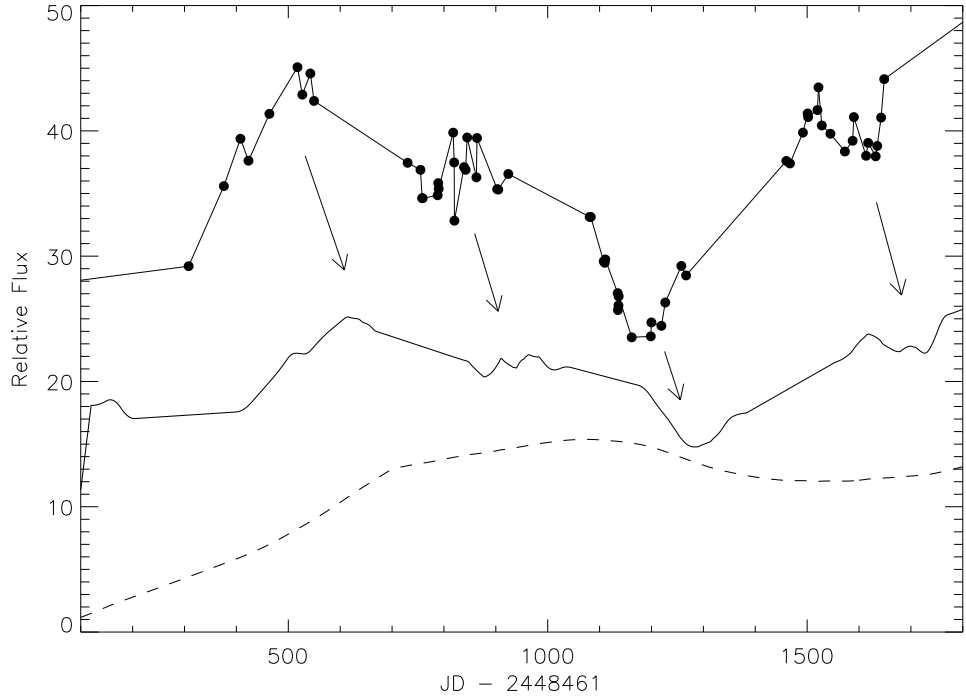}
\caption{Effects of extended size of BLR on observed light curve of broad emission line. 
Solid line plus solid circle represents the observed light curve of continuum 
emission of PG0052,  solid line represents the created light curve of emission line 
for the case that BLR with extended size about 30 light-days, dashed line represents 
the created light curve of emission line for the case that BLR with extended size 
about 600 light-days. The arrows show the positions of apparent features in light 
curve of continuum emission and in corresponding created light curve of emission line. 
}
\label{var}
\end{figure*}

\begin{figure*}
\centering\includegraphics[height = 10cm,width = 14cm]{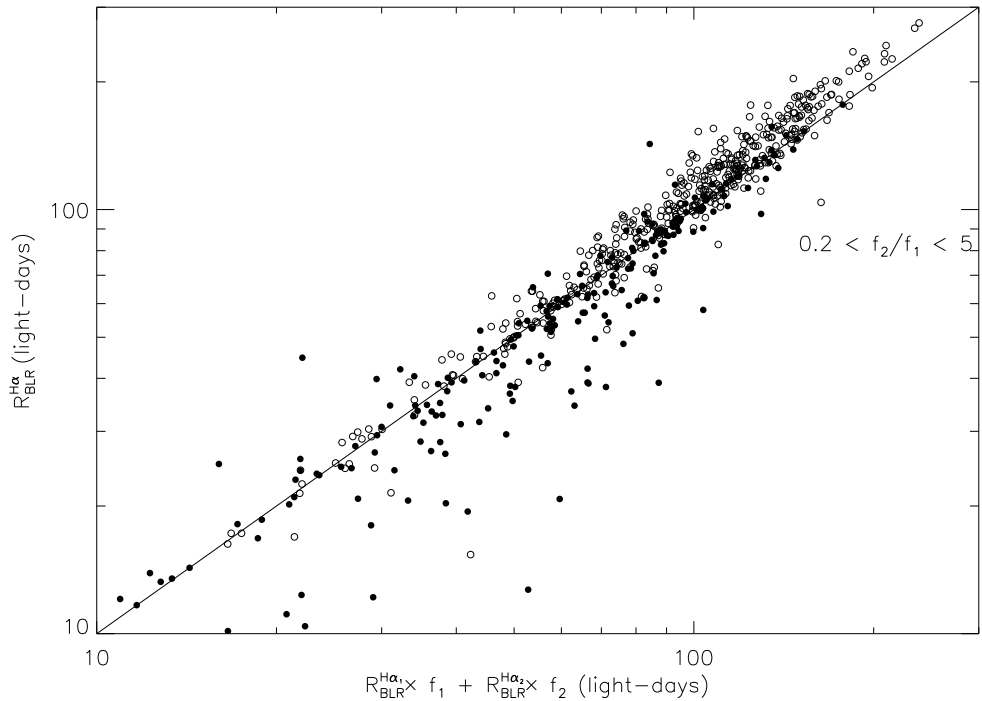}
\caption{On the correlation between $\RHa$ and
$\RHaa\times f_1 + \RHaaa\times f_2$ for total 664 simulated data points.
Open circles represent 444 data points with $f2/f1\ge1$, and
solid circles are for 220 data points with $f2/f1\le1$. Solid line
represents the relation
$R_{BLR}^{H\alpha}=\RHaa\times f_1 + \RHaaa\times f_2$.
}
\label{hsize}
\end{figure*}

\end{document}